\documentclass[fleqn,twoside]{article}
\usepackage{espcrc2}

\usepackage{graphicx}
\newcommand{\AmS}{{\protect\the\textfont2
  A\kern-.1667em\lower.5ex\hbox{M}\kern-.125emS}}

\title{Lattice-QCD based Schwinger-Dyson approach for Chiral Phase Transition}

\author{Hideaki~Iida,\hspace{-0.15cm}
\address{Dept. of Physics, Tokyo Institute of Technology, Ohokayama 2-12-1, 
Meguro, Tokyo 152-8551, Japan
}
Makoto~Oka$^{\rm a}$ and Hideo~Suganuma$^{\rm a}$
}

\begin{document}

\begin{abstract}
\vspace{-0.1cm}
Dynamical chiral-symmetry breaking in QCD is studied with 
the Schwinger-Dyson (SD) formalism based on lattice QCD data, i.e., LQCD-based SD formalism.
We extract the SD kernel function $K(p^2)$ in an Ansatz-independent manner 
from the lattice data of the quark propagator in the Landau gauge. 
As remarkable features, we find infrared vanishing and intermediate enhancement 
of the SD kernel function $K(p^2)$. 
We apply the LQCD-based SD equation to thermal QCD with the quark chemical potential $\mu_q$.
We find chiral symmetry restoration at $T_c \simeq 100{\rm MeV}$ for $\mu_q=0$. 
The real part of the quark mass function decreases as $T$ and $\mu_q$. 
At finite density, there appears the imaginary part of the quark mass function, 
which would lead to the width broadening of hadrons.
\vspace{-0.2cm}
\end{abstract}

\maketitle

\section{Introduction}

The Schwinger-Dyson (SD) approach~\cite{H84,M93,SST95,BPRT03,FA03} 
has been used as a nonperturbative calculation method for  
dynamical chiral-symmetry breaking (DCSB) in QCD.
The SD formalism consists of an infinite series of nonlinear integral equations 
which determine the $n$-point Green function of quarks and gluons.
Hence, it includes the infinite-order effect of the gauge coupling constant $g$.

The SD equation for the quark propagator $S(p)$ is described with 
the nonperturbative gluon propagator $D_{\mu\nu}(p)$ and 
the nonperturbative quark-gluon vertex $g \Gamma_\nu(p,q)$ as 
\begin{eqnarray}
S^{-1}(p)\!=\!S^{-1}_0(p) 
\!+\!g^2\!\! \int_q\!\! \gamma_\mu S(q) D_{\mu\nu}(p-q) \Gamma_\nu(p,q),
\label{eqn:SDE1}
\end{eqnarray}
where $S_0(p)$ denotes the bare quark propagator and 
the simple notation $\int_q \equiv \int\frac{d^4 q}{(2\pi)^4}$ has been used in the Euclidean metric.

In the practical calculation for QCD, however, the SD formalism is drastically truncated: 
the perturbative gluon propagator and the one-loop running coupling are used instead of 
the nonperturbative quantities in the original formalism.
By this simplification, some nonperturbative QCD effects will be neglected
and therefore the results may not be trustable.

In this paper, we formulate the SD equation based on the recent lattice QCD (LQCD) results, 
i.e., the LQCD-based SD equation \cite{IOS03}, and apply it to 
DCSB at finite temperatures and densities.

\section{The Quark Propagator in Lattice QCD}

In the Euclidean metric,  
the quark propagator $S(p)$ in the Landau gauge
is generally given by
\begin{eqnarray}
S(p)=Z(p^2)/\{\not p+M(p^2)\},
\label{eqn:quarkprop}
\end{eqnarray}
with the quark mass function $M(p^2)$ and the wave-function renormalization factor $Z(p^2)$. 

Recent quenched lattice QCD \cite{BHW02} indicates
\begin{eqnarray}
M(p^2)={M_0}/\{1+(p/\bar p)^\gamma\}
\label{eqn:quarkmass}
\end{eqnarray}
with $M_0\simeq$260 MeV, $\bar p\simeq$870MeV and $\gamma\simeq$3.04 
in the range of $0 \le p \le 4{\rm GeV}$ in the chiral limit.

The infrared quark mass $M(p^2=0)=M_0 \simeq 260{\rm MeV}$ seems consistent with the 
constituent quark mass in the quark model. 
Using this lattice result of $M(p^2)$, the pion decay constant is calculated as $f_\pi \simeq$ 87 MeV 
with the Pagels-Stokar formula, 
and the quark condensate is obtained as $\langle \bar qq \rangle_{\Lambda =1{\rm GeV}} \simeq -(220{\rm MeV})^3$.
These quantities on DCSB seem consistent with the standard values.

\section{The Schwinger-Dyson Equation}

In the Landau gauge, the Euclidean gluon propagator is generally expressed by 
\begin{eqnarray}
D_{\mu\nu}(p^2)= d(p^2)/p^2 \cdot ( \delta_{\mu \nu} - p_{\mu} p_{\nu}/p^2),
\label{eqn:gluonprop}
\end{eqnarray}
where we refer to $d(p^2)$ as the gluon polarization factor. 
For the quark-gluon vertex, 
we assume the chiral-preserving vector-type vertex, 
\begin{eqnarray}
\Gamma_\mu(p,q)=\gamma_\mu\Gamma((p-q)^2),
\label{eqn:vertex}
\end{eqnarray}  
which keeps the chiral symmetry properly.
(Recent lattice QCD studies indicate dominance of the vector vertex with  
 some nontrivial structure in the quark-gluon vertex \cite{SK02}.)
Taking the trace of Eq.(\ref{eqn:SDE1}), one gets 
\begin{eqnarray}
\frac{M(p^2)}{Z(p^2)}=3C_F\int_q  \frac{Z(q^2)M(q^2)}{q^2+M^2(q^2)} \frac{K((p-q)^2)} {(p-q)^2},
\label{eqn:SDE5}
\end{eqnarray}
where we define the kernel function
\begin{eqnarray}
K(p^2) \equiv g^2 \Gamma(p^2) d(p^2)
\label{eqn:kernel}
\end{eqnarray}
as the product of the quark-gluon vertex function $\Gamma(p^2)$ and the gluon polarization factor $d(p^2)$. The color factor of quarks is $C_F=4/3$.

\vspace{-0.1cm}

\section{Extraction of the SD Kernel Function}

In usual, the SD equation is used to obtain the quark mass function $M(p^2)$.
We note however that the SD equation is the relation between the quark propagator and $K(p^2)$.
Then, we extract the kernel function $K(p^2)$ 
in the SD equation (\ref{eqn:SDE5}) 
using the quark propagator obtained in lattice QCD. 
By shifting the integral variable from $q$ to $\tilde q \equiv p-q$, 
we obtain 
\begin{eqnarray}
\frac{M(p^2)}{Z(p^2)}=\frac{3C_F}{8\pi^3}\int_0^\infty d \tilde q^2 \Theta(p, \tilde q) K(\tilde q^2), 
\label{eqn:SDE8}
\end{eqnarray}
where $\Theta(p,q)$ is defined with  $M(p^2)$ and $Z(p^2)$ as 
\begin{eqnarray}
\Theta(p,q) \equiv \int_0^\pi d\theta \sin^2 \theta \hspace{3.5cm}
\nonumber \\ 
\frac{Z(p^2+q^2-2pq\cos\theta)M(p^2+q^2-2pq\cos\theta)}{p^2+q^2-2pq\cos\theta+M^2(p^2+q^2-2pq\cos\theta)}. 
\label{eqn:Theta}
\end{eqnarray}
From the knowledge of $M(p^2)$ and $Z(p^2)$ in lattice QCD \cite{BHW02}, 
we can calculate 
$K(p^2)$ with Eq.(\ref{eqn:SDE8}).

Figure 1 shows the kernel function $K(p^2)$ extracted from the quark propagator in lattice QCD.
\begin{figure}[t]
\centering
\rotatebox{-90}{\includegraphics[width=2in]{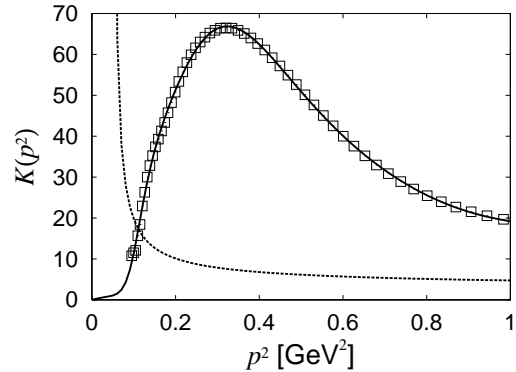}}
\vspace{-0.75cm}
\caption{
The SD kernel function $K(p^2)$
extracted from the quark propagator in lattice QCD in the Landau gauge. 
The square symbols are the calculated result, 
and the solid curve denotes a fit function for them.
The dotted curve corresponds to the ordinary SD kernel with the one-loop running coupling constant.
$K(p^2)$ exhibits infrared vanishing and intermediate enhancement.}
\vspace{-0.5cm}
\end{figure}
As remarkable features, 
we find ``infrared vanishing" and ``intermediate enhancement" 
in the SD kernel $K(p^2)$.
In fact, $K(p^2)$ seems consistent with zero in the very infrared region as 
$
K(p^2 \sim 0)\simeq 0, 
$
while $K(p^2)$ exhibits a large enhancement in the intermediate-energy region around $p \sim$ 0.6GeV.

Note that, to extract the kernel function $K(p^2)\equiv g^2 \Gamma(p^2)d(p^2)$, 
we use only the quark propagator and never use the gluon propagator.
Nevertheless, we find infrared vanishing and intermediate enhancement, which 
are also observed in 
the polarization factor $d(p^2)$ in the gluon propagator 
in the Landau gauge~\cite{L98}.
In fact, these two features of $K(p^2)$ are embedded in the information of the quark propagator in lattice QCD.

The dotted curve in Fig.1 corresponds to 
the usage of the perturbative gluon propagator $d(p^2)=1$ and the one-loop running coupling $g_{\rm run}(p^2)$, i.e., 
$K(p^2)=g^2_{\rm run}(p^2)$, which largely differs from the present result based on lattice QCD 
both in the infrared and in the intermediate-energy regions.
Hence, the simple version of the SD equation using 
the perturbative gluon propagator and the one-loop running coupling 
would be too crude for the quantitative study of QCD.

\vspace{-0.1cm}

\section{Chiral Symmetry at Finite Temperature}

We demonstrate a simple application of 
the LQCD-based SD equation to chiral symmetry restoration in finite-temperature QCD \cite{IOS03}. 

At a finite temperature $T$, 
the field variables obey the (anti-)periodic boundary condition 
in the imaginary-time direction, which leads to 
the SD equation for the thermal quark mass $M_n({\bf p}^2)$ 
of the Matsubara frequency $\omega_n \equiv (2n+1)\pi T$ and spatial three momentum ${\bf p}$ as 
\begin{eqnarray}
\hspace{-0.1cm}
\frac{M_n({\bf p}^2)}{Z_n({\bf p}^2)}
\hspace{-0.03cm}
=
\hspace{-0.07cm}
\int_{m,{\bf q}} 
\hspace{-0.06cm}
\frac{3C_F 
Z_m({\bf q}^2)M_m({\bf q}^2)}{\omega_m^2+{\bf q}^2+M_m^2({\bf q}^2)} 
\frac{K(\omega_{nm}^2
\hspace{-0.07cm}
+
\hspace{-0.02cm}
\tilde {\bf q}^2)}
{\omega_{nm}^2
+
\tilde {\bf q}^2}
\label{eqn:SDET}
\end{eqnarray}
with
$\int_{m,{\bf q}} \equiv T \sum_{m=-\infty}^\infty \int \frac{d^3q}{(2\pi)^3}$, 
$\omega_{nm} \equiv \omega_n-\omega_m$ and
$\tilde{\bf q} \equiv {\bf p}-{\bf q}$.

Using the kernel function $K(p^2)$ obtained in the previous section, 
we solve Eq.(\ref{eqn:SDET}) for the thermal quark mass $M_n({\bf p}^2)$.
Figure 2 shows the result for 
the thermal infrared quark mass $M_0({\bf p}^2=0)$ 
plotted against the temperature $T$.
We thus find chiral symmetry restoration at 
a critical temperature $T_c \simeq 100{\rm MeV}$.

\begin{figure}[hb]
\vspace{-0.9cm}
\centering
\rotatebox{-90}{\includegraphics[width=2in]{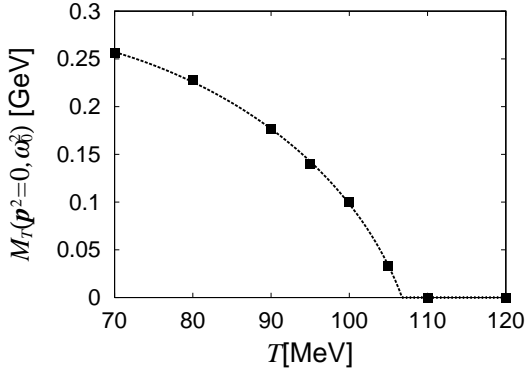}}
\vspace{-0.7cm}
\caption{
The thermal infrared quark mass $M_0({\bf p}^2=0)$ plotted against $T$ at zero density.
}
\vspace{-0.9cm}
\end{figure}

\section{Chiral Symmetry at Finite Density}
Finally, we apply the LQCD-based SD equation to thermal QCD with the quark chemical potential $\mu_q$, 
and investigate the density effect to the QCD phase transition. 
Since the quark wave-function renormalization effect seems small \cite{IOS03}, we set $Z(p^2)=1$ hereafter.

In finite density QCD, the quark chemical potential $\mu_q$ is introduced as
\begin{eqnarray}
H_{\rm QCD}(\mu_q)\equiv H_{\rm QCD}+\mu_q\hat{N}_q
\end{eqnarray}
with $\hat{N}_q\equiv \int d^3 x q^{\dagger}q$. This corresponds to 
\begin{eqnarray}
{\cal L}_{\rm QCD}(\mu_q)\equiv {\cal L}_{\rm QCD}-i\mu_q q^\dagger q
\end{eqnarray}
in Euclidean metric.
Accordingly, we make the replacement $\omega_n\rightarrow \omega_n-i\mu_q$ for the quark field. 

Note that there appears the imaginary part in the SD equation 
due to the explicit breaking of charge conjugation of the system. 
Then, we write the quark mass function $\tilde M_n({\bf p}^2)\in {\bf C}$ as
\begin{eqnarray}
\tilde{M}_n({\bf p}^2, \mu_q)\equiv M_n({\bf p}^2, \mu_q)+i\Gamma_n({\bf p}^2, \mu_q)
\end{eqnarray}
with real functions, $M_n({\bf p}^2, \mu_q)$ and $\Gamma_n({\bf p}^2, \mu_q)$.

Here, we expand the SD equation by the chemical potential $\mu_q$, 
and calculate the quark mass function order by order. 
As a merit of the expansion by $\mu_q$, the formalism is extremely simplified.
In principle, this expansion can be done to arbitrary order of $\mu_q$.
In general, the even-order contribution is real and the odd-order contribution is pure imaginary, 
because of the reflection symmetry of the SD equation in terms of $\mu_q\rightarrow -\mu_q$ and the complex conjugation, 
i.e., $\tilde{M}_n^*({\bf p}^2, -\mu_q)=\tilde{M}_n({\bf p}^2, \mu_q)$. 
Then, the quark mass function can be written as
\begin{eqnarray}
\hspace{-0.6cm} && M_n({\bf p}^2, \mu_q) \equiv M_n^{(0)}({\bf p}^2)+\hspace{-0.08cm}\mu_q^2M_n^{(2)}({\bf p}^2)+\cdots\\
\hspace{-0.6cm} && \Gamma_n({\bf p}^2, \mu_q) \equiv \mu_q\Gamma^{(1)}_n({\bf p}^2) + \mu_q^3\Gamma_n^{(3)}({\bf p}^2)+\cdots.
\end{eqnarray}

We calculate the quark mass function up to the second order of the quark chemical potential $\mu_q$. 
The 0th-order SD equation for $M_n^{(0)}({\bf p}^2)$ is 
the same as the SD equation (\ref{eqn:SDET}) at zero density.

The 1st-order SD equation for 
$\Gamma_n^{(1)}({\bf p}^2)$ reads
\begin{eqnarray}
\hspace{-0.7cm}&&\Gamma_n^{(1)}({\bf p}^2)=
\int_{m,{\bf q}}
\frac{K(\omega_{mn}^2,({\bf p}-{\bf q})^2)}{\omega_{mn}^2+({\bf p}-{\bf q})^2}\frac{3C_F}{D_m({\bf q}^2)^2}
\nonumber\\
\hspace{-0.7cm}
&&\{(\omega_m^2+{\bf q}^2-M_m^{(0)2}({\bf q}^2)) \Gamma_m^{(1)}({\bf q}^2)
+2\omega_m M_m^{(0)}({\bf q}^2)\}~~~~~~ \nonumber
\end{eqnarray}
with
$D_m({\bf q}^2) \equiv \omega_m^2+{\bf q}^2+{M_m^{(0) 2}({\bf q}^2})$.

The 2nd-order SD equation for 
$M_n^{(2)}({\bf p}^2)$ reads
\begin{eqnarray}
\hspace{-0.7cm}&&M_n^{(2)}({\bf p}^2)=
\int_{m,{\bf q}}
\frac{K(\omega_{mn}^2,({\bf p}-{\bf q})^2)}{\omega_{mn}^2+({\bf p}-{\bf q})^2}
\frac{3C_F}{D_m({\bf q}^2)^2} \nonumber\\
\hspace{-0.7cm}&&\{(\omega_m^2+{\bf q}^2-M_m^{(0)2}({\bf q}^2)) M_m^{(2)}({\bf q}^2)+C_m^{(2)}({\bf q}^2) \}.~~~~~~ \nonumber
\end{eqnarray}
with the non-homogeneous term, 
\begin{eqnarray}
\hspace{-0.7cm}&&C_m^{(2)}({\bf q}^2) \equiv
-\frac{4M_m^{(0)}({\bf q}^2)(M_m^{(0)}({\bf q}^2)\Gamma_m^{(1)}({\bf q}^2)-\omega_m)^2}
{D_m({\bf q}^2)}
 \nonumber\\
\hspace{-0.7cm}&&
+M_m^{(0)}({\bf q}^2)-2\omega_m\Gamma_m^{(1)}({\bf q}^2)
+3M_m^{(0)}({\bf q}^2)\Gamma^{(1)2}_m({\bf q}^2).~~~~~~ \nonumber
\end{eqnarray}

\begin{figure}[h]
\centering
\vspace{-0.15cm}
\rotatebox{-90}{\includegraphics[width=2in]{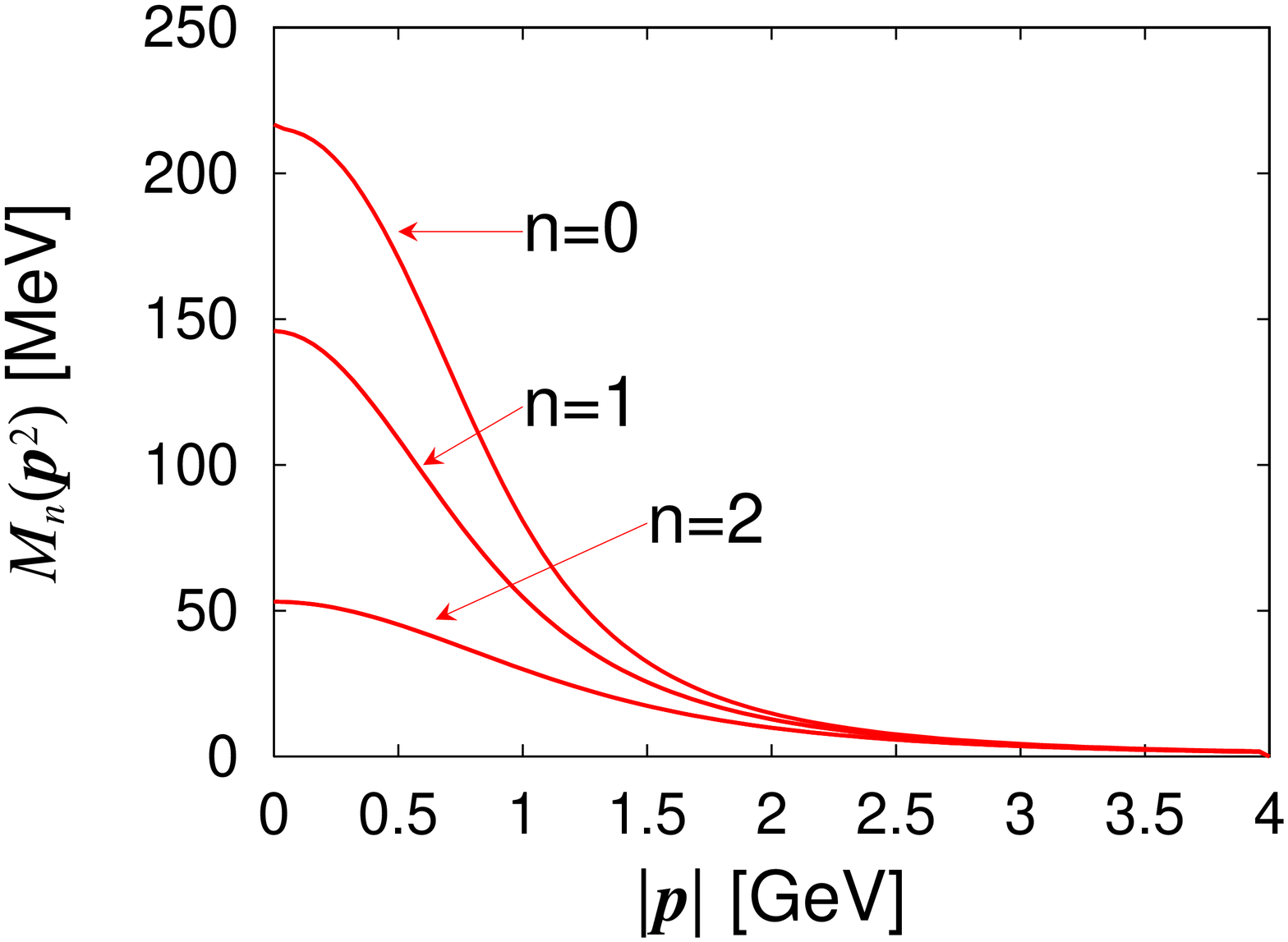}}
\vspace{-0.2cm}
\rotatebox{-90}{\includegraphics[width=2in]{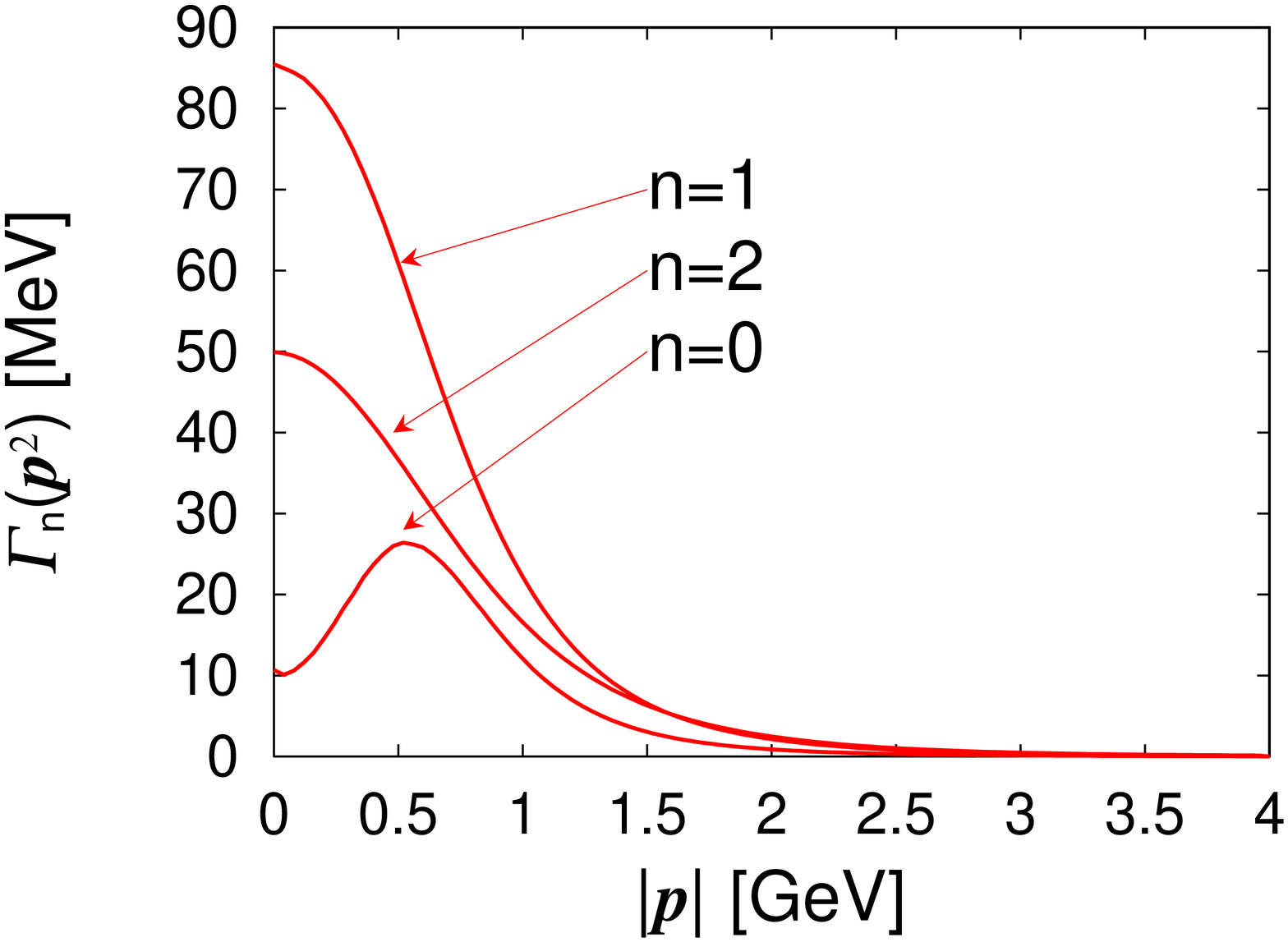}}
\vspace{-0.55cm}
\caption{
The real part of the quark mass function, 
$M_n({\bf p}^2, \mu_q) \equiv {\rm Re}  \tilde{M}_n({\bf p}^2, \mu_q)$ ($n$=0,1,2), 
and the imaginary part, 
$\Gamma_n({\bf p}^2, \mu_q) \equiv {\rm Im}  \tilde{M}_n({\bf p}^2, \mu_q)$ ($n$=0,1,2), 
plotted against the spatial momentum $|{\bf p}|$ for $T=70{\rm  MeV}$, $\mu_q=300{\rm MeV}$. 
}
\vspace{-0.73cm}
\end{figure}

Figure~3(a) shows the real part of the quark mass function, 
$M_n({\bf p}^2, \mu_q)={\rm Re}  \tilde{M}_n({\bf p}^2, \mu_q)$, 
for $T=70{\rm MeV}$ and $\mu_q=300{\rm MeV}$. 
$M_n({\bf p}^2)$ is found to decrease with $\mu_q$, which would indicate  
partial chiral symmetry restoration at finite density. 
Figure 3(b) shows the imaginary part of the quark mass function, 
$\Gamma_n({\bf p}^2, \mu_q)={\rm Im}  \tilde{M}_n({\bf p}^2, \mu_q)$.
As an interesting feature, $|\Gamma_0({\bf p}^2, \mu_q)|$ is smaller than $|\Gamma_1({\bf p}^2, \mu_q)|$, 
while $|\Gamma_n({\bf p}^2, \mu_q)|$ decreases with $n$ for $n \ge 1$ in this case of $T=70{\rm MeV}$.

As a physical consequence of the appearance of the imaginary part $\Gamma({\bf p}^2)$ in the quark mass function, 
all the composite particles of quarks, such as the $\rho$ meson and the nucleon, 
are expected to have an imaginary part of their spectral function at finite density 
in the constituent quark picture.
In fact, our result indicates the ``width broadening" of all the hadrons at finite density.

In 1995, the CERES Collaboration observed a large change of the vector meson spectrum 
in the heavy-ion collision experiment at the CERN SPS \cite{CERES95}. 
This change seems to be explained as the ``width broadening" rather than the ``mass shift" of the vector meson \cite{EIK98}. 
In this point, our result supports the width broadening together with a small mass shift 
at finite density 

\vspace{-0.1cm}

\section{Summary and Concluding Remarks}

We have studied DCSB in QCD with
the SD formalism based on lattice QCD data, i.e., LQCD-based SD formalism.
We have extracted the SD kernel function $K(p^2)$ in an Ansatz-independent manner 
from the lattice data of the quark propagator in the Landau gauge. 
We have found that the SD kernel $K(p^2)$ exhibits infrared vanishing and a large 
enhancement at the intermediate-energy region around $p\sim 0.6{\rm GeV}$.

We have applied the LQCD-based SD formalism to thermal and dense QCD.
We have found that the real part of the quark mass function decreases as $T$ and $\mu_q$.
At finite density, the quark mass function has an imaginary part, which would 
lead to the width broadening of hadrons.


\vspace{-0.1cm}

\begin{thebibliography}{9}
\bibitem{H84} K.~Higashijima, Phys. Rev. {\bf D29} (1984) 1228.
\bibitem{M93} V.~Miransky, Dynamical Symmetry Breaking in Quantum Field Theories, (WSPC, 1993).
\bibitem{SST95} H.~Suganuma, S.~Sasaki and H.~Toki, 
Nucl. Phys. {\bf B435} (1995) 207.
\bibitem{BPRT03} M.S.Bhagwat, M.A.Pichowsky, C.D.Roberts, P.C.Tandy, Phys. Rev. {\bf C 68} (2003) 015203. 
\bibitem{FA03} C.S.~Fischer and R.~Alkofer, Phys. Rev. {\bf D67} (2003) 094020; C.S.~Fischer, this proceedings.
\bibitem{IOS03} H.~Iida, M.~Oka and H.~Suganuma, Nucl. Phys. {\bf B} (Proc.Suppl.) {\bf 129-130} (2003) 602; 
hep-ph/0312328.
\bibitem{BHW02} P.O.~Bowman, U.M.~Heller, A.G.~Williams, Nucl. Phys. {\bf B} (Proc. Suppl.) {\bf 109} (2002) 163.
\bibitem{SK02} J.~Skullerud and A.~Kizilersu, JHEP {\bf 0209} (2002) 013; J.~Skullerud, this proceedings.
\bibitem{L98} UKQCD Collaboration (D.~Leinweber et al.), Phys. Rev. {\bf D58} (1998) 031501.
\bibitem{CERES95}
CERES Collaboration (G. Agakishiev et al.), Phys. Rev. Lett. {\bf 75} (1995) 1272.
\bibitem{EIK98} V. Eletsky, B. Ioffe, J. Kapusta, Eur.Phys.J. {\bf A3} (1998) 381; 
Nucl.Phys. {\bf A661} (1999) 514.
\end{thebibliography}
\end{document}